\documentclass[showpacs,preprintnumbers,amsmath,amssymb,nofootinbib]{revtex4}
\usepackage{graphicx}
\usepackage{epsfig}
\usepackage{bm}
\usepackage{amsfonts}

\newcommand{\eq}{\begin{eqnarray}}
\newcommand{\eqx}{\end{eqnarray}}
\newcommand{\ba}{\begin{equation}}
\newcommand{\ea}{\end{equation}}
\newcommand{\f}[2]{\frac{#1}{#2}}

\newcommand{\ep}{\epsilon}
\newcommand{\bit}{\begin{itemize}}
\newcommand{\eit}{\end{itemize}}

\def\la{\label}

\def\bi{\bibitem}

\def\g{\gamma}
\def\d{\partial}
\def\th{\theta}

\def\om{\omega}
\begin{document}

\title{Entropy flow of a perfect fluid in (1+1) hydrodynamics}

\author{Guillaume Beuf}
\email{guillaume.beuf@cea.fr} \affiliation{Institut de Physique Th\'eorique,\\
CEA, IPhT, F-91191 Gif-sur-Yvette, France\\
CNRS, URA 2306 }

\author{Robi Peschanski }
\email{robi.peschanski@cea.fr} \affiliation{Institut de Physique Th\'eorique,\\
CEA, IPhT, F-91191 Gif-sur-Yvette, France\\
CNRS, URA 2306 }

\author{Emmanuel N.~Saridakis }
\email{msaridak@phys.uoa.gr} \affiliation{Department of Physics,\\
University of Athens,\\ GR-15771 Athens, Greece}

\begin{abstract}
Using the formalism of the Khalatnikov potential, we derive
exact general formulae for the entropy flow dS/dy, where y is the
rapidity, as a function of temperature for the (1+1) relativistic
hydrodynamics of a perfect fluid. We study in particular flows dominated by a
sufficiently long hydrodynamic evolution, and provide an
explicit analytical solution for dS/dy. We discuss the theoretical
implications of our general formulae and some phenomenological applications for
heavy-ion collisions.
\end{abstract}

\pacs{12.38.Mh,24.10.Nz} \maketitle

\section{Introduction}

There is an accumulating evidence that hydrodynamics may be
relevant for the description of the medium created in high-energy
heavy ion collisions \cite{revhyd}. Indeed, experimental
measurements such as the elliptic flow \cite{flow} shows the
existence of a collective effect on the produced particles which
can be described in terms of a motion of the fluid. More
precisely, numerical simulations of the hydrodynamic equations
describe quite well the distribution of low-$p_\perp$ particles
\cite{revhyd}, with an equation of state close to that of a
``perfect fluid'' with a rather low viscosity. On the other hand,
it seems useful  to discuss a simplified picture \cite{lan,bj},
which can be qualitatively understood in physical terms, namely
the idea that the evolution of the system before freeze-out is
dominated by the longitudinal motion. Thus, the hydrodynamic
transverse motion can be neglected or at least factorized out in
order to study the longitudinal flow only.

Indeed, the two seminal   applications of relativistic
hydrodynamics to particle and heavy-ion collisions, by Landau
\cite{lan} and by Bjorken \cite{bj}, start with this $1\!+\!1$
approximation, valid in the determinative stage of the reaction.
The longitudinal hydrodynamic approach has found many
applications. It has been used in the literature
\cite{(1+1),Carruthers} in order to discuss aspects of the
hydrodynamical flow which are relevant for the physical
understanding of  high-energy particle scattering and, more
recently \cite{stein}, of heavy-ion collisions.

Soon after the first proposal by Landau and its derivation of a
large-time approximation \cite{lan}, Khalatnikov \cite{khal}
showed that $(1\!+\!1)$ hydrodynamics derive from a potential
verifying a  linear equation. The Khalatnikov potential has been
used in the literature in an initial period \cite{bi,old}, but has
not been recently considered, to our knowledge. Very recently,
the interest on looking for exact solutions of (1+1) hydrodynamics
has been revived and one finds new examples and applications of
exact solutions, {\it e.g.} \cite{Bipe,new}. For instance, in a
recent paper \cite{Bipe}, a unified description of Bjorken and
Landau (1+1) flows has been proposed as a class of exact solutions
of (1+1) hydrodynamics based on harmonic flows. (1+1)
hydrodynamics appears also quite recently in the application of
string-theoretical ideas to the formation a strongly interacting
quark-gluon plasma \cite{janik}. These exact solutions allow to
find explicit analytical solutions for relevant observables. Among
these, the entropy flow $dS/dy$, where y is the rapidity, is quite
interesting, since it may be related to the multiplicity
distribution of particles. Our goal is to go beyond particular
cases and obtain a general expression of the entropy flows as a
function of the temperature for a generic solution of (1+1)
hydrodynamics, i.e. for a generic solution of the Khalatnikov
equation.

We are interested in the distribution $dS/dy$ of entropy  density
per unit of the rapidity $y$ which is related to  the flow
velocities $u_\pm =e^{\pm y}$ in the (1+1) approximation.  This
``hydrodynamic observable'' depends in an essential way on the
assumed hypersurface\footnote{We keep the term {\it hypersurface}
in the (1+1) case to keep track, even formally,  of the transverse
motion.} through which we want to compute the flow (and eventually
relate it to physical observables in collisions). For given
entropy $s$ and energy $\epsilon$ densities, $dS/dy$  depends on
the hypersurface one considers to follow the hydrodynamic
evolution. It is particularly interesting  to consider
hypersurfaces corresponding to a fixed-temperature. Indeed, it
allows to follow the cooling of the hydrodynamic flow, from an
initial stage characterized by a high temperature, towards a final
stage which is often associated  to a freeze-out temperature.
Hence our aim is to derive an expression of $dS/dy$ as a function
of temperature for a given Khalatnikov potential and to
investigate its properties, both on theoretical and
phenomenological points of view.

Our main new result, $i.e.$ the general expressions for the
entropy flow as a function of the Khalatnikov potential, can be
found in three different versions (\ref{dS_FO}, \ref{dS_FO_bis},
\ref{dS_FO_ter}). One or the other expressions can be more
suitable for a given explicit problem.

The plan of our paper is as follows. First, in section
\ref{model}, we group, for completion, all the necessary material,
including  the hydrodynamic equations,  the Khalatnikov potential,
its equation and solutions, recast and derived in a modern
framework using light-cone variables. In section \ref{entrflow},
we formulate and derive the general expression of the entropy flow
in rapidity as a function of
the temperature evolution. In section \ref{general}, we derive and
study a family of exact solutions, namely the ones where the final
entropy distribution is dominated by the hydrodynamical evolution
and not by the initial conditions. They generalize the Landau
flow, and  give the asymptotic behavior of physical flows in the
limit of long hydrodynamical evolution. We provide in particular
the exact analytic expression of the final entropy distribution
corresponding to the Belenkij-Landau \cite{bi} solution. Then, in section
\ref{pheno}, we compare
the profile of the entropy distributions, as well as their energy
dependence, with the relevant experimental data. The final section
is traditionally devoted to conclusions and outlook.

\section{(1+1) Relativistic hydrodynamics of a perfect fluid }
\label{model}
\subsection{Hydrodynamic equations}

We consider a perfect fluid whose  energy-momentum tensor is
\begin{equation}
 T^{\mu\nu}= (\epsilon+p)u^{\mu}u^{\nu} - p \eta^{\mu\nu}  \label{Tmn}
\end{equation}
 where $\epsilon$ is the energy density, $p$ is the
pressure and $u^{\mu}$ ($\mu =\{0,1,2,3\}$) is the  4-velocity in the
Minkowski metric $\eta^{\mu\nu}$. It obeys the  equation
\begin{equation}
\d_\mu T^{\mu\nu}=0\ .
 \la{hydro}\end{equation}
Using the standard thermodynamical identities  (where we have
assumed for simplicity vanishing chemical potential):
\begin{equation}
p+\epsilon = Ts\;;\;\; d\epsilon = T ds\;;\;\; dp = s dT
\la{therm}\ ,
\end{equation}
the system of hydrodynamic equations closes  by relating energy
density and pressure through the general equation of state
\begin{equation}
\frac{dp}{d\ep}=\frac{s dT}{T ds}=c_s^2(T)\ . \label{egpvar}
\end{equation}
We consider now the (1+1) approximation of the hydrodynamic flow,
restricting it only to  the longitudinal direction. Within such an
approximation, the effect of the transverse dimensions is only
reflected through the equation of state \eqref{egpvar}. Note that
we do not $a \ priori$ assume the traceless condition $T^{\mu\mu}=0,$
and thus the fluid is considered as ``perfect'' (null viscosity)
but not necessarily ``conformal'' (null trace).

Let us introduce  light-cone coordinates
\begin{equation}
 z^\pm= t\pm z= z^0\pm z^1 =\tau e^{\pm \eta}\;
\Rightarrow \; \left(\f {\d}{\d z^0}\pm \f {\d}{\d z^1}\right)=2
\f {\d}{\d z^\pm} \equiv2 \d_\pm \label{dz}
 \end{equation} where
$\tau=\sqrt{z^+z^-}$ is the proper time and
$\eta=\frac{1}{2}\ln({z^+}/{z^-})$ is the space-time rapidity of
the fluid. We also introduce for further use the light-cone
components of the fluid velocity
 \ba u^\pm\equiv u^0\pm u^1=e^{\pm y}\ ,\label{uplus}\ea
 where $y$ is the usual rapidity variable (in the energy-momentum space).

The hydrodynamic equations (\ref{hydro}) take the form
\begin{eqnarray}
 \left(\partial_++
\partial_-\right)T^{00}+\left(\partial_+- \partial_-\right)T^{01}&=&0
\nonumber \\
\left(\partial_++ \partial_-\right)T^{01}+\left(\partial_+-
\partial_-\right)T^{11}&=&0\ .
\label{eqbasic}
\end{eqnarray}

\subsection{Khalatnikov potential}
\label{Khalderivation}

It is known \cite{khal,bi} that one can
replace to the non-linear problem of (1+1) hydrodynamic
evolution with a linear equation for a suitably defined potential.
In this section we follow the method of Ref.\cite{khal},  recasting the
calculations
in the light-cone variables.

Inserting in (\ref{eqbasic}) the known relations (\ref{Tmn})  for
$T^{\mu\nu}$and expressing everything in light-cone coordinates
using (\ref{dz},\ref{uplus}), one
obtains the following two equations:
\begin{eqnarray}
\left(\frac{e^{2y}-1}{2}\right)\partial_+\left(\ep+p\right)+
e^{2y}\left(\ep+p\right)\partial_+y+\left(\frac{1-e^{-2y}}{2}\right)\partial_-
\left(\ep+p\right)+
e^{-2y}\left(\ep+p\right)\partial_-y+\partial_+p-\partial_-p&=&0
\nonumber \\
\left(\frac{e^{2y}+1}{2}\right)\partial_+\left(\ep+p\right)+
e^{2y}\left(\ep+p\right)\partial_+y+\left(\frac{1+e^{-2y}}{2}\right)\partial_-
\left(\ep+p\right)-
e^{-2y}\left(\ep+p\right)\partial_-y-\partial_+p-\partial_-p&=&0\
.\label{eqbasicvar00}
\end{eqnarray}
In (\ref{eqbasicvar00})  the energy density $\ep$ and pressure $p$
are considered as functions of the kinematic light-cone variables
$(z^+,z^-)$. One key ingredient of the potential method
\cite{khal} is to express the  hydrodynamic equations in terms of
the hydrodynamical  variables $y =  \log u^+= -\log u^-$ and
$\th=\log \left[{T}/{T_0}\right],$ where $T_0$ is an arbitrary
temperature scale.

Relations (\ref{eqbasicvar00}) can be further transformed by
inserting the differentials of the thermodynamic relations
(\ref{therm}), namely
\begin{eqnarray}
\partial_\pm\left(\ep+p\right)=T_0\ \partial_\pm\left(s\ e^\theta\right)
\nonumber \\
\partial_\pm p=T_0\ s\ \partial_\pm e^\theta\ . \label{epderiv}
\end{eqnarray}
Multiplying the first equation of (\ref{eqbasicvar00}) by
$(e^{-2y}+1)$, the second by $(e^{-2y}-1),$  adding them and using
(\ref{epderiv}), one obtains:
\begin{equation}
\partial_+\left(e^{\th+ y}\right)=\partial_-\left(e^{\th- y}\right)
.\label{rel1}
\end{equation}
Eq.\eqref{rel1} proves the existence of a potential\footnote{The function $\Phi$
has some
degree of  arbitrariness since we could define
$\partial_\mp\Phi\equiv T_0 e^{\th\pm y}+\varphi_\mp(z^\mp)$, with
$\varphi_-(z^-)$ and $\varphi_+(z^+)$ arbitrary
 one-variable functions. This freedom, analogous to a gauge choice, does not
modify  the final results.} $\Phi(z^+,z^-)$ verifying:
\begin{equation}
\partial_\mp\Phi(z^+,z^-)\equiv u^\pm T=T_0\ e^{\th\pm y}\ .\label{partialphi}
\end{equation}
In this way, (\ref{rel1}) is automatically satisfied.

In order to transform the  system of equations \eqref{eqbasicvar00} from the
kinematic
variables ($z^+,z^-$) to the dynamical ones ($\theta,y$), one
introduces \cite{khal} the Khalatnikov  potential $\chi$, considered as a
function of $(u^+T,u^-T)$ through a
Legendre transform:
\begin{equation}
\chi(u^+T,u^-T)\equiv\Phi(z^+,z^-)-z^-u^+T-z^+u^-T\ ,\label{chi0}
\end{equation}
where $z^{\pm}$ are functions of
$(u^+T,u^-T)$ implicitly defined by \eqref{partialphi}. Hence, we get:
\begin{equation}
\frac{\partial \chi}{\partial (u^\mp T)}\ =-z^{\pm}+\left[\partial_+
\Phi-u^-T\right]\ \frac{\partial z^+}{\partial (u^\mp T)}
+\left[\partial_- \Phi-u^+T\right]\frac{\partial z^-}{\partial
(u^\mp T)} \equiv-z^{\pm} \, ,\label{chiuTz}
\end{equation}
where, due to the relations (\ref{partialphi}), the terms between
brackets are  zero. Knowing the Khalatnikov potential $\chi$, which is a
function of the thermodynamic variables, one can find the kinematic variables of
the flow by derivation.

In the following, we will always consider the Khalatnikov potential $\chi$ as
function of $\theta$ and $y$, keeping the same notation for $\chi$. That change
of variables corresponds for the differential operators to
\begin{equation}
\frac{\partial }{\partial (u^\pm T)}= \frac{1}{2 T_0}\,
\,e^{-\th \mp y}\, (\partial_\theta \pm\partial_y) \, .\label{derivCV}
\end{equation}
In those variables, relation \eqref{chiuTz} writes
\begin{equation}
z^\pm(\theta, y)=\frac{1}{2 T_0}\ e^{-\theta \pm y}\ \left(-\partial_\theta\chi
\pm \partial_y\chi\right)\, . \label{zpzm}
\end{equation}
From \eqref{zpzm}, one also gets the expressions for the proper time $\tau$ and
the space-time rapidity $\eta$ (defined as in \eqref{dz})
\begin{eqnarray}
\tau (\theta, y) &=& \frac{e^{-\theta}}{2 T_0} \ \sqrt{(\partial_\theta
\chi)^2-(\partial_y \chi)^2} \nonumber\\
\eta (\theta, y)&=&\ y+\frac{1}{2} \log
\left(\frac{-\partial_\theta \chi+\partial_y
\chi}{-\partial_\theta \chi-\partial_y \chi} \right) = y - \tanh^{-1}
\left(\frac{\partial_y
\chi}{\partial_\theta \chi}\right)\  \,
.\label{TauEta}
\end{eqnarray}

\subsection{Khalatnikov equation}
\label{Khalequation} Coming back to the system of equations
(\ref{eqbasicvar00}), another independent  combination can be
obtained. Multiplying  the first equation by $(e^{-2y}-1)$, the
second by $(e^{-2y}+1)$  and adding, we obtain
$\partial_+\left(e^{y}s\right)+\partial_-\left(e^{-y}s\right)=0$,
or equivalently:
\begin{equation}
\partial_+\left(u^+s\right)+\partial_-\left(u^-s\right)=0
\label{varentropeq}.
\end{equation}
This relation corresponds physically to the conservation of
the entropy  along the flow. It is a property of the perfect fluid
 that the motion of the pieces of the fluid along the velocity lines
is isentropic.

Following the logics of the Legendre transform, we transform  relation
(\ref{varentropeq}) using the
$(\theta,y)$-base. For this sake, we write down the following
partial derivatives:
\begin{eqnarray}
\frac{\partial(u^\pm s)}{\partial\theta}\equiv
u^\pm\frac{ds}{d\theta}&=&\frac{\partial(u^\pm s)}{\partial
z^+}\frac{\partial z^+}{\partial \theta}+\frac{\partial(u^\pm
s)}{\partial
z^-}\frac{\partial z^-}{\partial \theta}\nonumber\\
\frac{\partial(u^\pm s)}{\partial y}\equiv\pm u^\pm
s&=&\frac{\partial(u^\pm s)}{\partial z^+}\frac{\partial
z^+}{\partial y}+\frac{\partial(u^\pm s)}{\partial
z^-}\frac{\partial z^-}{\partial y}
 \label{Sdiff}.
\end{eqnarray}
Solving this system of linear equations  we obtain:
\begin{eqnarray}
\frac{\partial(u^+s)}{\partial z^+}&=&\
\frac{1}{D}\left[u^+\frac{\partial z^-}{\partial
y}\frac{ds}{d\theta}-u^+s\frac{\partial z^-}{\partial
\theta}\right]\nonumber\\
\frac{\partial(u^-s)}{\partial
z^-}&=&-\frac{1}{D}\left[u^-\frac{\partial z^+}{\partial
y}\frac{ds}{d\theta}+u^-s\frac{\partial z^+}{\partial
\theta}\right]
 \label{Sdiff2},
\end{eqnarray}
where\footnote{We  assume that the determinant $D$ is
non-zero, which is the case except on exceptional lines
\cite{bi}.}
\begin{equation}
D=\frac{\partial z^+}{\partial \theta}\frac{\partial z^-}{\partial
y}-\frac{\partial z^+}{\partial y}\frac{\partial z^-}{\partial
\theta}\ .\label{determ}
\end{equation}
Inserting (\ref{Sdiff2}) into the entropy-flow conservation
relation (\ref{varentropeq}) we acquire:
\begin{equation}
\frac{ds}{d\theta}\left[u^+\frac{\partial z^-}{\partial
y}-u^-\frac{\partial z^+}{\partial y}\right]
-s\left[u^+\frac{\partial z^-}{\partial \theta}+u^-\frac{\partial
z^+}{\partial \theta}\right]=0.\label{intermediate}
\end{equation}
Obtaining the expression of the $z^\pm$ derivatives from
(\ref{zpzm}), the equation (\ref{intermediate}) leads to:
\begin{equation}
\frac{1}{s}\frac{ds}{d\theta}\left[\partial_\theta
\chi-\partial^2_y \chi\right]-\partial_\theta \chi
+\partial^2_\theta \chi=0\ . \label{kal1}
\end{equation}
Making use of the sound velocity relation (\ref{egpvar}) we
finally arrive at the  Khalatnikov equation \cite{khal,bi}:
\begin{equation}
c^2_s\,\partial_\theta^2
\chi(\theta,y)+\left[1-c_s^2\right]\partial_\theta
\chi(\theta,y)-\partial^2_y \chi(\theta,y)=0.
 \label{Khalatnikov}
\end{equation}
Hence, the non-linear system of equations which governs the (1+1)
hydrodynamical flow has been converted into a linear,
second-order, hyperbolic partial differential equation. Note that the
Khalatnikov equation is valid independently from the specific form
of the sound velocity.

\subsection{Application: solutions of the Khalatnikov equation for fixed $c_s$.}
\label{Khalsolution}

In this section, for our purpose, we present the solutions of the
Khalatnikov equation with a constant speed of sound:
\begin{equation}
c_s^2 \equiv \frac{p}{\ep}={g}^{-1}\ , \label{egpbis}
\end{equation}
where $g$ will be considered as  a parameter  in the
Khalatnikov equation (\ref{Khalatnikov}).  Note that in this  case
the general relations (\ref{therm}) write
\begin{equation}
\epsilon =gp=\epsilon_0\ \left(\frac{T}{T_0}\right)^{g+1}=\
\epsilon_0\ e^{(g+1)\theta} , \label{tempbis}\end{equation} for
the energy density and
\begin{equation}
s=s_0\ \left(\frac{T}{T_0}\right)^g =s_0\ e^{g\theta}
\label{entropybis}
\end{equation}
for the entropy density.

Writing
\begin{equation}
\chi(\theta,y)=e^{-\left(\frac{g-1}{2}\right) \theta}\ Z(\theta,y)
\label{Zvarbis}
\end{equation}
and inserting it into (\ref{Khalatnikov}), we acquire:
\begin{equation}
\partial^2_\theta Z-g\ \partial^2_yZ-{\scriptstyle \left(\frac{g-1}2\right)^2}\
Z=0,\label{eqnz}
\end{equation}
where we have used  a compact notation for  partial derivatives.

It is convenient to replace the variables $\theta$ and $y$ by
$\alpha$ and $\beta$, defined by
\begin{equation}
\alpha \equiv -\theta + \frac{y}{\sqrt{g}} \quad \textrm{and}
\quad \beta \equiv -\theta - \frac{y}{\sqrt{g}}\,
,\label{alphabeta}
\end{equation}
such that equation \eqref{eqnz} takes the form
\begin{equation}
\partial_\alpha \partial_\beta \bar{Z}(\alpha,\beta)-{\scriptstyle
\frac{(g-1)^2}{16}}\ \bar{Z}(\alpha,\beta)=0\ .\label{eqnzalphabeta}
\end{equation}
We solve  this equation following the Green's functions formalism,
i.e. we look for distributions $\bar{G}(\alpha,\beta)$ such that
\begin{equation}
\partial_\alpha \partial_\beta \bar{G}(\alpha,\beta)-{\scriptstyle
\frac{(g-1)^2}{16}}\
 \bar{G}(\alpha,\beta)=\delta(\alpha) \delta(\beta)\, .\label{eqGreenalphabeta}
\end{equation}
The relevant solution of equation \eqref{eqGreenalphabeta}
is\footnote{There exist other Green's functions of equation
\eqref{eqGreenalphabeta}, with \emph{e.g.} $\Theta(-\alpha)$
instead of $\Theta(\alpha)$, or $\Theta(-\beta)$ instead of
$\Theta(\beta)$. Assuming that the fluid naturally expands and
cools down during the evolution, and taking the arbitrary
temperature scale $T_0$ to be the maximal temperature of the
sources, $-\theta$ increases with time. Thus
\eqref{solGreenalphabeta} gives the only physical solution of
equation \eqref{eqGreenalphabeta}, analogous to the retarded
propagator of the D'Alembert equation. Finally, note that for the
obvious physical requirement of  finite behavior at
$\alpha,\beta\rightarrow\infty$,  we reject the solutions of
\eqref{eqGreenalphabeta} containing the Bessel-$K_0$ function
instead of $I_0$.}
\begin{equation}
\bar{G}(\alpha,\beta)= \Theta(\alpha)\ \Theta(\beta)\
I_0\left(\frac{g-1}{2}\sqrt{\alpha \beta} \right)\,
,\label{solGreenalphabeta}
\end{equation}
with $I_0$ the modified Bessel function of the first kind and
$\Theta$ the Heaviside function. Using the relation
\begin{equation}
\delta(\alpha)\,
\delta(\beta)\equiv\delta\left(-\theta+\frac{y}{\sqrt{g}}\right)\,
\delta\left(-\theta-\frac{y}{\sqrt{g}}\right)= \sqrt{g}\
\delta(\theta)\ \delta(y)\, ,
\end{equation}
we deduce from \eqref{solGreenalphabeta} the relevant Green's
function of \eqref{eqnz}:
\begin{equation}
G(\theta,y)=\frac{1}{4\sqrt{g}}\
\bar{G}(\alpha,\beta)=\frac{1}{4\sqrt{g}}\,
\Theta\left(-\theta+\frac{y}{\sqrt{g}}\right)\,
\Theta\left(-\theta-\frac{y}{\sqrt{g}}\right)\,
I_0\left(\frac{g-1}{2}\sqrt{\theta^2-\frac{y^2}{g}}
\right)\label{solGreenThetay},
\end{equation}
which verifies
\begin{equation}
\partial^2_\theta G-g\ \partial^2_y G-{\scriptstyle \left(\frac{g-1}2\right)^2}\
G=\delta(\theta) \delta(y)\, .\label{eqGreenthy}
\end{equation}
Thus, we can construct the general solution of Khalatnikov
equation \eqref{Khalatnikov}, inserting a distribution of sources
$F(\hat{\theta},\hat{y})$, as:
\begin{eqnarray}
\chi(\theta,y)&=&e^{-\left(\frac{g-1}{2}\right) \theta}\ \int
d\hat{y}\
\int d\hat{\theta}\, G(\theta-\hat{\theta},y-\hat{y})\,
F(\hat{\theta},\hat{y})\nonumber\\
&=& \frac{e^{-\left(\frac{g-1}{2}\right) \theta}}{4\sqrt{g}}\ \int
d\hat{y}\ \int_{\theta+|y-\hat{y}|/\sqrt{g}}^{+\infty}
d\hat{\theta}\, F(\hat{\theta},\hat{y}) \,
I_0\left(\frac{g-1}{2}\sqrt{(\theta-\hat{\theta})^2-\frac{(y-\hat{y})^2}{g}}
\right)\, .\label{solGenKhalat}
\end{eqnarray}

Equation (\ref{solGenKhalat}) gives the most general solution, for
any distribution of sources of hydrodynamic flow. In the context
of heavy-ion collisions, we are mostly interested in solutions
that correspond to the evolution of a flow starting from initial
conditions on a curve of the $(\theta,y)$ plane. Therefore, one
should impose constraints on $F(\hat{\theta},\hat{y})$, in order
to decribe the initial conditions. In section \ref{general} we
will consider a physically-interesting sub-class of solutions.

\section{Derivation of the entropy flow}
\label{entrflow}

Coming back to the general formalism, let us now derive  the exact formula for
the entropy flow $dS/dy$
at a given fixed temperature $T_F= T_0\ e^{\th_F} $, as a function
of rapidity $y.$ For a general (1+1) hydrodynamic expansion we
consider the solution  formulated in terms of the general
Khalatnikov potential  $\chi(\th,y),$ given by
(\ref{solGenKhalat}). The entropy distribution at  fixed
temperature is expressed through the amount of entropy flowing
through the hypersurface of fixed temperature $T_F,$ in an
infinitesimal rapidity interval. It is given by (see \emph{e.g.} \cite{Bipe})
\begin{equation}
\frac{dS}{dy}\equiv s_F \frac{u^\mu d\lambda_\mu}{dy} = s_F u^\mu
n_\mu \frac{d\lambda}{dy}\, ,\label{dsdy1}
\end{equation}
where $d\lambda$ is the infinitesimal (space-like) length element
along the hypersurface of fixed temperature $T_F,$ and $n^{\mu}$
is the normal to the hypersurface. The entropy density  depends
only on the temperature and not on $y.$ Hence it is constant along
the fixed-temperature hypersurface, namely $s_F\equiv
s(T_F)\propto T_F^g\ $.

\subsection{The flow through the fixed-temperature hypersurface}

As we have mentioned, we concentrate on hypersurfaces at fixed
temperature $T_F$ (or equivalently at $\theta_F = \log
\left[{T_F}/{T_0}\right]$). It is convenient to use as
 kinematical functions the proper time $\tau=\sqrt{z^+z^-}$  and the
space-time rapidity $\eta=\frac{1}{2}\ln({z^+}/{z^-})$, considered
as functions of $\theta$ and $y$. In this ($\theta,y$)-base, the
fixed-temperature hypersurface is parameterized by
\begin{eqnarray}
\tau_F(y) &=& \tau(\theta_F,y)\nonumber\\
\eta_F(y) &=& \eta(\theta_F,y)\label{paramFOut}
\end{eqnarray}
considered as functions of $y$ at $\th_F$ fixed. The tangent
vector to the hypersurface reads:
\begin{eqnarray}
V^+ (y)\equiv z_F^{+'}(y) &=& (\tau_F' + \eta_F' \tau_F)\ e^{\eta_F}\nonumber\\
V^- (y)\equiv z_F^{-'}(y) &=& (\tau_F' - \eta_F' \tau_F)\
e^{-\eta_F} ,\label{tangvect}
\end{eqnarray}
where the primes denote derivatives with respect to $y$. Hence, we
can construct the normalized perpendicular vector to the
fixed-temperature curve $(n^+(y),n^-(y))$ defined by
\begin{eqnarray}
n^+ (y)\ n^- (y) &=& 1\nonumber\\
\frac{1}{2}\ [n^+ (y)\ V^- (y) +  n^- (y)\ V^+ (y)]&=& 0 \ .
\end{eqnarray}
Using (\ref{tangvect}) the second equation translates into
\begin{equation}
n^+ (y)\ e^{-\eta_F}\ (\eta_F' \tau_F - \tau_F') = n^- (y)\
e^{\eta_F}\ (\eta_F' \tau_F + \tau_F')\, .
\end{equation}
Provide $|\eta_F'(y)| > \left|\frac{\tau_F'(y)}{\tau_F(y)}
\right|$ for all $y$, we find
\begin{eqnarray}
n^+ (y)&=& \sqrt{\frac{\eta_F' \tau_F + \tau_F'}{\eta_F' \tau_F - \tau_F'}} \
e^{\eta_F}\nonumber\\
n^- (y)&=& \sqrt{\frac{\eta_F' \tau_F - \tau_F'}{\eta_F' \tau_F +
 \tau_F'}}\ e^{-\eta_F}.\label{nplus}
\end{eqnarray}
Following Ref.\cite{Bipe}, $d \lambda^{\mu} \equiv d \lambda \
n^{\mu}$ is defined  such that
\begin{equation}
(d \lambda)^2 = d \lambda^{\mu} d \lambda_{\mu} = - dz_F^+ \
dz_F^- = -(\tau_F'^2 - \tau_F^2\ \eta_F'^2) (dy)^2 ,
\end{equation}
where the minus sign comes from the fact that the hypersurface is
a space-like curve. Thus, we have
\begin{equation}
d \lambda = \sqrt{\tau_F^2\ \eta_F'^2 - \tau_F'^2} \ dy\, .
\label{dsigma}
\end{equation}

So, inserting (\ref{nplus}) and (\ref{dsigma}) in (\ref{dsdy1}),
we finally find
\begin{equation}
\frac{dS}{dy}(y) = s_F \left[\tau_F(y)\ \eta_F'(y)\ \cosh
(\eta_F(y) -y) + \tau_F'(y)\ \sinh (\eta_F(y)
-y)\right].\label{help1}
\end{equation}

\subsection{Expression of the entropy flow}

Let us now introduce the expression of  the entropy flow in terms
of the Khalatnikov potential. Starting from (\ref{zpzm}), we  obtain:
\begin{eqnarray}
\cosh (\eta-y) &=& -\frac{1}{2 \tau T_0 e^\theta}\ \partial_\theta \chi
(\theta,y)\nonumber\\
\sinh (\eta-y) &=& \frac{1}{2 \tau T_0 e^\theta}\ \partial_y \chi (\theta,y)\,
.\label{NeumannCond}
\end{eqnarray}

Now, inserting (\ref{NeumannCond}) in (\ref{help1}) we can eliminate
the hyperbolic trigonometrical functions acquiring:
\begin{equation}
\frac{dS}{dy}(y) = s_F\left.\frac{\left[- \tau_F(y)\ \eta_F'(y)\
\partial_\theta \chi(\theta,y) + \tau_F'(y)\ \partial_y
\chi(\theta,y)\right]}{2T_0\ e^{\theta_F}\
\tau_F(y)}\right|_{\theta=\theta_F}.\label{help2}
\end{equation}
Furthermore, by differentiation of the relations (\ref{TauEta}) with respect to
$y$,
at $\theta=\theta_F$, we find
\begin{eqnarray}
\tau_F'(y) &=& \frac{e^{-\theta}}{2}
\left.\frac{\left[(\partial_\theta \chi)
 (\partial_y \partial_\theta \chi)- (\partial_y \chi)
 (\partial_y^2 \chi)\right]}{\sqrt{(\partial_\theta \chi)^2-(\partial_y
\chi)^2}}\right|_{\theta=\theta_F}\nonumber\\
\eta_F'(y) &=& \left.\frac{\left[(\partial_\theta
\chi)^2-(\partial_y \chi)^2 + (\partial_y \chi)(\partial_y
\partial_\theta \chi) - (\partial_\theta \chi)(\partial_y^2 \chi)
\right]}{(\partial_\theta \chi)^2-(\partial_y
\chi)^2}\right|_{\theta=\theta_F}.\label{TauEtaDerivFO}
\end{eqnarray}

Then, inserting the relations \eqref{TauEtaDerivFO} and \eqref{TauEta} in
(\ref{help2}), we obtain a
remarkably simple expression, namely:
\begin{equation}
\frac{dS}{dy}(y) = \frac{s_F}{2 T_F}\,
\left.[\partial_y^2 \chi (\theta,y)-\partial_\theta
\chi(\theta,y)]\right|_{\theta=\theta_F} ,\label{dS_FO}
\end{equation}
which possesses a full  generality, as long as the Khalatnikov
potential $\chi(\th,y)$ exists. In addition, using the Khalatnikov
equation (\ref{Khalatnikov}), (\ref{dS_FO}) can be also written
as:
\begin{equation}
\frac{dS}{dy}(y) = \frac{s_F \ c_s^2(T_F)}{2 T_F}\,
\left.[\partial_\theta^2 \chi (\theta,y)-\partial_\theta
\chi(\theta,y)]\right|_{\theta=\theta_F} .\label{dS_FO_bis}
\end{equation}

There is an interesting  third version of equations
(\ref{dS_FO},\ref{dS_FO_bis}), featuring the potential $\Phi$ instead of $\chi$.
The definition \eqref{chi0} of $\chi$ can be written alternatively
\begin{equation}
\chi=\Phi-T \tau e^{\eta-y}-T \tau e^{-\eta+y}\equiv\Phi-2 T \tau \cosh
(\eta-y)\, .\label{chi0bis}
\end{equation}
Inserting in \eqref{chi0bis} the first relation of \eqref{NeumannCond}, one
obtains
\begin{equation}
\Phi=\chi(\theta,y)-\partial_\theta \chi(\theta,y)\, .\label{PhiChi}
\end{equation}
Inserting that last relation into \eqref{dS_FO_bis}, and considering the
potential $\Phi$ (originally defined in \eqref{partialphi} as a function of
$z^+$ and $z^-$) now as a function of $\theta$ and $y$, one gets a third
equivalent formula for the entropy flow through fixed-temperature hypersurfaces,
namely
\begin{equation}
\frac{dS}{dy}(y) = -\frac{s_F \ c_s^2(T_F)}{2 T_F}\,
\left.\partial_\theta \Phi_{} (\theta,y)\right\vert_{_{\theta=\theta_F}}
.\label{dS_FO_ter}
\end{equation}

The set of expressions (\ref{dS_FO}), (\ref{dS_FO_bis}) and
\eqref{dS_FO_ter} form our main formal result. They provide the
exact form of the entropy flow along fixed-temperature
hypersurfaces, for a general (1+1) hydrodynamic evolution. We also
mention  that, beyond the derivation of the Khalatnikov  potential
at fixed sound velocity, formulae
(\ref{dS_FO},\ref{dS_FO_bis},\ref{dS_FO_ter}) still hold for a
general speed of sound, once the solution of the general
Khalatnikov equation (\ref{Khalatnikov}) is known. It is important
to note that relations (\ref{dS_FO},\ref{PhiChi}) are valid as
long as there exist $\chi$ or $\Phi$ potentials, even if there is
no reduction to a linear equation, $i.e.$ no entropy conservation
in the (1+1) projection of the flow, while relations
(\ref{dS_FO_bis},\ref{dS_FO_ter}) are valid when the Khalatnikov
equation holds $i.e.$ with entropy conservation in the (1+1)
projection of the flow.

\subsection{Examples}

Let us check the general formulae for the entropy flow  considering  exact
hydrodynamical solutions known in the literature, namely the Bjorken flow
\cite{bj} and the harmonic flows \cite{Bipe}.\\

\paragraph{Bjorken flow:}

The Bjorken flow corresponds to boost-invariance, \emph{i.e.} $\partial_y
\chi\equiv 0$. In this case, the Khalatnikov equation \eqref{Khalatnikov}
reduces to
\begin{equation}
\chi''(\theta)+(g-1) \chi'(\theta)=0\, ,\label{KhalatEqBjorken}
\end{equation}
which has as generic solution
\begin{equation}
\chi(\theta)= C e^{-(g-1)\theta}\, ,\label{chiBjorkenBis}
\end{equation}
$C$ being an integration constant. Let us choose $C$ and the arbitrary
temperature scale $T_0$ such that at the proper time $\tau=\tau_0$, the
temperature of the fluid is $T=T_0$. Inserting \eqref{chiBjorkenBis} in
relations \eqref{TauEta} one  finds $C=2 T_0 \tau_0 /(g-1)$, and the known
expressions for the Bjorken flow, namely
\begin{equation}
\tau(\theta,y)=\tau_0\ e^{-g \theta} \qquad \textrm{and} \qquad \eta\equiv y\, ,
\label{BjorkenSol}
\end{equation}
\emph{i.e.} the equality of rapidity with space-time rapidity.
Finally, the Khalatnikov potential for the Bjorken solution writes
\begin{equation}
\chi(\theta)= \frac{2 T_0 \tau_0}{(g-1)} e^{-(g-1)\theta}=\frac{2 T
\tau}{(g-1)}\, .\label{chiBjorken}
\end{equation}
Inserting \eqref{chiBjorken} into \eqref{dS_FO}, one obtains the entropy flow
\begin{equation}
\frac{dS}{dy}(y) = s_F \tau_F = s_0 \tau_0 =cst.\, .
\end{equation}
Hence, as expected from boost invariance of the Bjorken flow, not only the total
entropy but also the entropy flow is conserved.\\

\paragraph{Harmonic flows:}

Following Ref.\cite{Bipe}, one is led to introduce new auxiliary variables
$l^+(z^+)$ and $l^-(z^-)$ satisfying
\begin{equation}
\frac{d l^\pm}{d z^\pm}= \lambda \ e^{-{l^\pm}^2} \label{lpmConstraint}\ ,
\end{equation}
where $\lambda=cst.$  The thermodynamic variables can be
explicitly written \cite{Bipe} as\footnote{Here we use our
convention $\theta = \log T/T_0$ with opposite sign w.r.t.
\cite{Bipe}.}
\begin{eqnarray}
\theta &=& -\frac{g+1}{4g}\left({l^+}^2+{l^-}^2\right)+\frac{g-1}{2g}\ l^+
l^-\nonumber\\
y &=& \frac{1}{2} \left({l^+}^2-{l^-}^2\right)\ .\label{thetaHarm}
\end{eqnarray}
Using  the property \eqref{partialphi} of the potential $\Phi$ one writes
\begin{equation}
\frac{\partial \Phi}{\partial l^\pm} = \frac{d z^\pm}{d l^\pm} \ \partial_\pm
\Phi = \frac{d z^\pm}{d l^\pm}\  T_0 \ e^{\theta \mp y}\ .
\label{1}\end{equation}
Now, inserting \eqref{lpmConstraint} and the expressions \eqref{thetaHarm}, one
obtains
\begin{eqnarray}
\frac{\partial \Phi}{\partial l^\pm} &=& \lambda  T_0\ e^{{l^\pm}^2} e^{\theta
\mp y}=  \lambda  T_0\ e^{\frac{g-1}{4g} (l^+ + l^-)^2}\ .\label{2}
\end{eqnarray}
The expressions \eqref{2} are symmetric in $l^+$ and $l^-$ and thus, by mere
integration, one gets
\begin{equation}
\Phi(l^+,l^-)=\lambda  T_0\ \int^{l^+ + l^-} \! dv\ e^{\frac{g-1}{4g} v^2}\
,\label{PhiHarm}
\end{equation}
where the potential can be expressed in terms  of $\theta$ and $y$
through\begin{equation} l^+ + l^- = \sqrt{2}\ |y|\
\left(-\theta-\sqrt{\theta^2-\frac{y^2}{g}}\right)^{-1/2}\label{lpluslminus}\!
.
\end{equation}
Now, using our relation \eqref{dS_FO_ter}, and the relation
\begin{equation}
\partial_\theta \Phi = \lambda T_0\ e^{\frac{g-1}{4g} (l^+ + l^-)^2}\
\partial_\theta (l^+ + l^-)\ ,
\end{equation}
one gets the result for the entropy flow
\begin{equation}
\frac{dS}{dy}(y) = \frac{\sqrt{2}\lambda T_0 s_F}{ g T_F}\ e^{\frac{g-1}2
\left(\theta+\sqrt{\theta^2-y^2/g}\right)}\ \frac{|y|}{\sqrt{\theta^2-y^2/g}}
\left(-\theta-\sqrt{\theta^2-y^2/g}\right)^{-1/2}\! .\label{dS_harmonic}
\end{equation}
Using our general formalism, we thus recover the nontrivial result
obtained by direct calculation (see Ref.\cite{Bipe}, formula
(58)). Interestingly enough, we note that for the family of
harmonic flows as an example, it appears to be much simpler to use
formula \eqref{dS_FO_ter} for the potential $\Phi$ than using the
Khalatnikov potential $\chi$ itself.

Note that
we have to make a specific discussion of the limiting case when $g=1,$ 
that is when the speed of sound equals the speed of light. In fact in this case, 
the harmonic flow cannot be obtained as above and the solution for the flow 
acquires a more general form. Coming back to equation \eqref{Khalatnikov}, one 
finds that the Khalatnikov potential itself is harmonic, namely $\chi(\theta,y) 
\equiv h_+(y\!+\!\theta\sqrt{g}) + h_-(y\!-\!\theta\sqrt{g})$ where $h_+,h_-$ 
are arbitrary functions. We thus recover the results noted in  Refs.\cite{g1}.

\section{Evolution dominated solutions} \label{general}

In general, a longitudinal flow in the final state follows from a
longitudinal pressure gradient and/or from a longitudinal flow in
the initial state. Let us consider the sub-class of solutions where
the effect of the initial flow is negligible compared to the one
of the initial pressure gradient. This sub-class corresponds to
the dominance of the hydrodynamic evolution over the influence of the initial
conditions. A typical example of such a solution is the Belenkij-Landau
solution \cite{bi}, where the fluid is initially at rest (the
so-called ``full stopping'' initial conditions), and then expands into
the vacuum.

\subsection{Khalatnikov potential and entropy flow}

In order to model an evolution-dominated flow,  let us consider
all the sources at rest, \emph{i.e.}
$F(\hat{\theta},\hat{y})\propto \delta (\hat{y})$. Let us also
take the arbitrary temperature scale $T_0$ to be the maximal
temperature of the sources (hence $\theta\equiv\log (T/T_0) \leq
0$), \emph{i.e.} $F(\hat{\theta},\hat{y})\propto
\Theta(-\hat{\theta})$. All in all, we write
\begin{equation}
F(\hat{\theta},\hat{y}) = 4 \sqrt{g}\ K(\hat{\theta})\
\Theta(-\hat{\theta})\ \delta(\hat{y})\,
.\label{LandauLikeSources}
\end{equation}
Inserting \eqref{LandauLikeSources} in \eqref{solGenKhalat}, and
replacing the  variable $\hat{\theta}$ by $\theta'\equiv
\theta-\hat{\theta}$, one gets
\begin{eqnarray}
\chi(\theta,y)=
e^{-\left(\frac{g-1}{2}\right)\theta}\int^{-\frac{|y|}{\sqrt{g}}}_{\theta}\
I_0\left(\frac{g\!-\!1}{2}\sqrt{\theta'^2\!-\!y^2/g}\right)
\,K(\theta\!-\!\theta')d\theta'\, ,\label{chi}
\end{eqnarray}
where the function $K(\theta\!-\!\theta')$ carries the information
on the initial conditions. Note that $\theta'$ is also
negative.

In the following it is convenient to use a Laplace representation
of \eqref{chi}. Since $\theta\leq0$, we introduce the Laplace
transform, and its inverse, with respect to $-\theta$ as:
\begin{eqnarray}
\tilde{f}(\gamma)=\int_{-\infty}^0d\theta\
e^{\gamma\theta}\ f(\theta)\nonumber\\
f(\theta)=\int^{\gamma_0+i\infty}_{\gamma_0-i\infty}\frac{d\gamma}{2\pi
i}\ e^{-\gamma\theta}\ \tilde{f}(\gamma)\
,\label{Laplacetransform}
\end{eqnarray}
where $\gamma_0$ is a real constant that exceeds the real part of
all the singularities of the integrand, i.e the integral is
calculated on an imaginary  contour that lies on the right of all
singularities. Following \cite{bi},  the
Khalatnikov potential \eqref{chi} can be written  as a convolution
of the two functions :
\begin{eqnarray}
\Theta(-\theta)\,K(\theta)&=&\int^{\gamma_0+i\infty}_{\gamma_0-i\infty}
\frac{d\gamma}{2\pi
i}\,e^{-\gamma\theta}\tilde{K}(\gamma)\nonumber\\
\Theta\left(-\!\theta\!-\!|y|/\sqrt{g}\right)I_0\left(\frac{g\!-\!1}{2}
\sqrt{\theta^2\!-\!y^2/g}\right)&=&
\int^{\gamma_0+i\infty}_{\gamma_0-i\infty}\frac{d\gamma}{2\pi
i}\,\frac{1}{\sqrt{\gamma^2-\frac{(g-1)^2}{4}}}\left[e^{-\gamma\theta-\frac{|y|}
{\sqrt{g}}\sqrt{\gamma^2-\frac{(g-1)^2}{4}}}\right]
\, .\ \ \ \ \ \ \ \ \ \ \label{convgen}
\end{eqnarray}
As the Laplace transform changes convolutions into ordinary
products, one gets the Laplace representation
\begin{equation}
\chi(\theta,y)=\int^{\gamma_0+i\infty}_{\gamma_0-i\infty}\frac{d\gamma}{2\pi
i}\left[e^{-\left(\gamma+\frac{g-1}{2}\right)\
\theta-\frac{|y|}{\sqrt{g}}\sqrt{\gamma^2-\frac{(g-1)^2}{4}}}\right]\,
\frac{\tilde{K}(\gamma)}{\sqrt{\gamma^2-\frac{(g-1)^2}{4}}}\
.\label{chiktild}
\end{equation}

Notice that, while the expression of the
solution (\ref{chi}) restricts the phase-space  domain in the
interval $\vert y\vert \le -\sqrt g\ \theta$, equation
(\ref{chiktild}) may allow for an analytic continuation of the
solution of the Khalatnikov potential outside this region. However the outside
region may be  different
($e.g.$ with $\chi \equiv 0$, as in \cite{bi}).

Let us now investigate the properties of the entropy flow given by
the solutions  \eqref{chiktild} of the Khalatnikov  equation.
Inserting the Khalatnikov potential (\ref{chiktild}) into the
expression of the entropy distribution (\ref{dS_FO_bis}), one is
led to the following formula:
\begin{eqnarray}
\!\!\!\frac{dS}{dy}(y)=\frac{s_F}{2g
T_F}\int^{\gamma_0+i\infty}_{\gamma_0-i\infty}\frac{d\gamma}
{2\pi i}\
e^{-\theta_F(\gamma+\frac{g-1}{2})}\left[\left(\gamma\!+\!
g/2\right)^2\!\!-\!\!1/4\right]\  {\tilde{K}(\gamma)}\ \frac
{e^{{-\frac{|y|}{\sqrt{g}}\sqrt{\gamma^2-
\frac{(g-1)^2}{4}}}}}{\sqrt{\gamma^2-\frac{(g-1)^2}{4}}}\ \,\
.\label{genentr}
\end{eqnarray}
In formula (\ref{genentr}), one may distinguish the {\it kernel}
\ba {\cal Q}(\g,y)\equiv \frac
{\exp{{-\frac{|y|}{\sqrt{g}}\sqrt{\gamma^2-
\frac{(g-1)^2}{4}}}}}{\sqrt{\gamma^2-\frac{(g-1)^2}{4}}}
\la{kernel} \ea
 driving the dynamical hydrodynamic evolution as  expressed on
  the entropy flow, and the {\it coefficient function}
\ba
\Tilde{C}_f(\gamma)=\left[\left(\gamma\!+\!g/2\right)^2\!-\!1/4\right]\
{\tilde{K}(\gamma)} \la{coeff} \ea
which encodes the initial conditions of the entropy flow.

\subsection{Total entropy}

Since we have a well-defined relation \eqref{genentr} for the
entropy distribution, it is easy to perform the integration over
$y$ and obtain the total entropy flux through the hypersurface
with fixed temperature  $T=T_F$.

Formally, (\ref{genentr}) leads to:
\begin{eqnarray}
S_{tot}\left|_{\theta=\theta_F}\right.\ \ &=&
2\int_0^{-\theta_F\sqrt{g}}\! dy\
\frac{s_F}{2g
T_F}\,\int^{\gamma_0+i\infty}_{\gamma_0-i\infty}\frac{d\gamma}{2\pi
i}\,\left(\gamma+\frac{g\!-\!1}{2}\right)\left(\gamma+\frac{g\!+\!1}{2}\right)\,
e^{-\frac{|y|}{\sqrt{g}}\sqrt{\gamma^2-\frac{(g-1)^2}{4}}}\ \times \nonumber\\
& & \qquad \qquad \qquad \qquad  \qquad \qquad  \qquad \qquad \qquad
\qquad\times \frac{\tilde{K}(\gamma)}{\sqrt{\gamma^2-\frac{(g-1)^2}{4}}}
\,e^{-\theta_F(\gamma+\frac{g-1}{2})}\label{genentr2}
\ ,\label{GentotS3}
\end{eqnarray}
where we took into account the $\Theta\left(-\!\theta\!-\!|y|/\sqrt{g}\right)$ 
function present in \eqref{convgen}. Indeed, the hydrodynamical flow is limited 
in the region inside this domain, with possible contributions on the boundary 
$\!\theta\!= -\!|y|/\sqrt{g}$ (Riemann waves, see, $e.g.$ \cite{bi}).

We know that, by construction,  the 
flow is isentropic and thus the total 
entropy $S_{tot}$ is conserved. In fact it is possible to show that the dominant 
part of the total conserved entropy results from the kernel \eqref{kernel} more 
than other sources such as the coefficient function \eqref{coeff} or the 
boundary Riemann waves. Hence the hydrodynamical dynamics dominate.
For this sake, let us release for simplicity the boudary limitations of the 
integral over $y.$ One writes
\begin{equation}
S_{tot} \approx \frac{s_F}{T_F
\sqrt{g}}\,\int^{\gamma_0+i\infty}_{\gamma_0-i\infty}\frac{d\gamma}{2\pi
i}\,\frac{\gamma+\left(\frac{g+1}{2}\right)}{\gamma-\left(\frac{g-1}{2}\right)}
\,\tilde{K}(\gamma)\,e^{-\theta_F(\gamma+\frac{g-1}{2})}
%\left\{1-e^{-\frac{Y_{max}}{2\sqrt{g}}\sqrt{\gamma^2-\frac{(g-1)^2}{4}}}\right}
= \frac{s_F\sqrt{g}}{T_F}\,\tilde{K}\left[(g\!-\!1)/{2}\right]\
e^{-(g-1)\theta_F}\label{GentotS4}\ .
\end{equation}
Indeed,  the complex integral is obtained through
 the singularities of the integrand, which can be due either to
the initial conditions (through singularities of
$\tilde{K}(\gamma)$) or to the hydrodynamical dynamics (through
the pole at $\gamma=(g-1)/{2}$), or both. If the singularities of
$\tilde{K}(\gamma)$ are situated at the left (resp. right) of the
pole, they will be subdominant (resp. dominant) in the total
entropy. Assuming a dominance of the hydrodynamic flow, we get
the final result 
of  (\ref{GentotS4}).
The physical meaning of  (\ref{GentotS4}) becomes
clear when using the thermodynamical relation
(\ref{entropybis}) and the entropy density $s_0$ at the
temperature $T_0.$ The total entropy writes\footnote{Note that $\tilde{K}$
is dimensionless as the potential $\chi$.}
\begin{equation}
S_{tot} \approx \frac{s_0\sqrt{g}}{T_0}\,\tilde{K}\left[(g\!-\!1)/{2}\right]\
\label{GentotS5},
\end{equation}
and thus does not depend on the features of the flow at $T=T_F$.
In conformity with the isentropic property of the flow, the total
hydrodynamic entropy  of the perfect fluid should be  conserved, as the
expression \eqref{GentotS5} is independent of $T_F$. This provides
a self-consistency check for an evolution-dominated flow.
In more general cases, 
one should also take into account the other contributions.

A final comment is in order. A priori,
the domain of integration $|y| \le Y/2$ comes from energy-momentum
conservation. However, the formula (\ref{chi}) for the Khalatnikov
potential is  only valid in the domain $|y| \le
-\sqrt g\ \th_F$.
In the flow-dominated approximation,
 $-\sqrt g\th_F$ and
$Y/2$ are considered large enough such that the integration domain can extend
 to infinity and the kernel only contributes\footnote{We have performed 
numerical checks
which show that thanks to the decreasing exponential behavior, the  boundary
term contributes
negligibly to the total entropy.}.

\subsection{Relation to the Belenkij-Landau solution}

We have studied the dependence of our results on the coefficient
function (\ref{coeff}) by imposing various relevant analytic forms
for $\tilde{K}(\gamma)$. We observed that typical meromorphic
functions bounded by a constant\footnote{Indeed, choosing
$\tilde{K}(\gamma)$ of strictly positive degree leads to an
unphysical angular point at $y=0$ and to a function $K(\theta)$,
see (\ref{convgen}), containing derivatives of Dirac
distributions, i.e. structures that are too singular to describe
physical flows.} at infinity and with poles at the left of
$\gamma=\frac{g-1}{2}$, give smooth and similar entropy flow
distributions, almost identical at large enough $\theta_F$. Hence
we conjecture that all physical evolution-dominated solutions to
be almost identical, at least for a sufficiently large value of
$\theta_F=\log (T_F/T_0)$. In order to provide an analytic expression for
the entropy flow characteristic of the family of solutions, we
remark that the following choice of the coefficient function
(\ref{coeff})
\begin{equation}
 \Tilde{C}_f(\gamma)=C \left(\gamma
+\frac{g-1}{2}\right)\ \ \  \Leftrightarrow\ \ \
\tilde{K}(\gamma)=\frac{C}{(\gamma+\frac{g+1}{2})},\label{lanbil}
\end{equation}
where C is a dimensionless constant, corresponds to the
hydrodynamical flow with an initial full stopping
condition \cite{bi}.

The Belenkij-Landau solution \cite{bi} describes the evolution of a slice of 
fluid of width $2L$ initially at rest and expanding in the vacuum. It consists 
in a hydrodynamical flow bounded by Riemann waves. The matching conditions 
between the flow and the waves in space-time translated in terms of temperature 
and rapidity variables are realized by imposing zero boundary conditions on the 
Khalatnikov potential $\chi$ on the characteristics $\theta=\pm y /\sqrt{g}$. 
Another condition on the potential is that the center of the slice remains by 
symmetry at rest ($y=0$) during the evolution.

We have checked
that the energy flow, resulting from modifications of the Ansatz
\eqref{lanbil} satisfying the dominance of the kernel singularity, is not
sensibly modified from the one given by inserting the coefficient
function (\ref{lanbil}) into (\ref{genentr}).

Inserting now (\ref{lanbil}) into (\ref{chiktild}), the
Khalatnikov potential between the characteristics $-\theta\ge |y| /\sqrt{g}$ 
acquires the analytic form \cite{bi,old} 
\begin{equation}
\chi(\theta,y)=C\,\int^{-\frac{|y|}{\sqrt{g}}}_{\theta}
I_0\left(\frac{g\!-\!1}{2}\sqrt{\theta'^2-\frac{y^2}{g}}\right)
\,e^{\th-\left(\frac{g+1}{2}\right)\theta'}d\theta' \
.\label{chilan}
\end{equation}
The potential is identically zero in the region $-\theta\le |y| /\sqrt{g}$.
Note that the constant in (\ref{lanbil},\ref{chilan}) is such that $C\propto L 
T_0$ with our notations.

Let us now insert this specific solution to our general formula
(\ref{dS_FO_bis}) for the entropy distribution. Calculating the
derivatives we find:
\begin{equation}
\frac{dS}{dy}(y) = s_F \frac{(g\!-\!1) C}{4g\ T_F}
\,e^{-\frac{(g-1)}{2}\theta_F}
\left[I_0\left(\frac{g\!-\!1}{2}\sqrt{\theta_F^2-y^2/g}\right)-
I_1\left(\frac{g\!-\!1}{2}\sqrt{\theta_F^2-y^2/g}\right)
\frac{\theta_F}{\sqrt{\theta_F^2-y^2/g}}\right]\label{Lansol3}\ .
\end{equation}
Since $\theta_F$ is negative, this expression is always positive,
at least in the region $\theta_F^2-y^2/g \ge 0.$  Hence the
positivity of the entropy flow is ensured. Finally, the expression
(\ref{Lansol3}) is divergence-free, since it is finite for
$\theta_F^2=y^2/g$. We also note that it is still real  in the
analytic continuation of the solution (\ref{Lansol3}) for
$\theta_F^2<y^2/g$, since in this case both the numerator and the
denominator of the second term are purely
imaginary\footnote{Positivity may also extend, but is not ensured
due to  the appearance of  Bessel function zeroes.}.

For the total entropy, inserting (\ref{lanbil}) into the general
formula (\ref{GentotS3}) one gets, using the thermodynamic relations
\eqref{entropybis}
\begin{equation}
S_{tot} =\frac{C\ s_F}{\sqrt{g}\ T_F}\ e^{-(g-1)\theta_F}
 = \frac{C\ s_0}{\sqrt{g}\ T_0}
\label{GentotSLan}\ .
\end{equation}

A comment is in order at this point. Condition (\ref{lanbil}) has
been considered  to describe the so-called full stopping
conditions. In the original papers \cite{bi}, it consists in the
assumptions that i) there is a specific plane where the medium is
at rest for all times, and that ii) on the vacuum-boundary we have
just a simple (Riemann) wave. In fact, the resulting entropy flow
distribution is expected to be more general and is characteristic
of the evolution-dominated hydrodynamic flows. Hence the
Khalatnikov potential (\ref{chilan}) (already obtained in
\cite{bi}) and the entropy flow (\ref{Lansol3}) may serve as an
analytic formulation for the class of evolution-dominated flows.
In fact, their features are essentially determined by the
evolution kernel ${\cal Q}(\g,y)$ (see formula (\ref{kernel})).

Finally, it is interesting to note that formula (\ref{genentr})
gives the possibility to  compare the hydrodynamic predictions
with those of other existing models of heavy-ion (and eventually
hadron-hadron, soft scattering) reactions. This relies on the
possibility of relating thermodynamic quantities, such as the
temperature and the entropy, to observed properties of the
particle multiplicities. In this scheme, the
rapidity of particles is defined by the corresponding value of
$y\equiv \log u^+$ obtained from  the fluid velocity of the lump
of fluid giving rise locally to the hadrons. In the same context,
the overall temperature gradient $\th_F$ will be related to the
total available rapidity and the entropy to the
multiplicity up to phenomenological factors. We will discuss in
the next section  the phenomenological issues of our derivation,
but the general theoretical idea is that formula  (\ref{genentr})
can be compared with the one-particle inclusive hadronic cross section which is
related to the scattering
amplitudes. In this respect the generic form (\ref{kernel}) of the
hydrodynamic evolution kernel may serve for a comparison of hydrodynamic
properties with conventional models of scattering amplitudes.

\section{Phenomenological applications}
\la{pheno}

Motivated by the seminal works of Landau \cite{lan} and Bjorken
\cite{bj}, the comparison of their predictions for (1+1)
hydrodynamics with some features of the data has been made (see \emph{e.g.}
\cite{bi,old,(1+1),Carruthers,stein}). Even
if such a rough  approximation, ignoring the details of the
transverse motion  or of  the hadronization, cannot replace the numerical
simulations, it has given some useful information on the dynamics
of the quark-gluon plasma. For instance, the order of magnitude
estimates made using the Bjorken flow in the central region
\cite{bj} and the comparison of the multiplicity distributions
with the predictions of the Landau flow \cite{Carruthers,stein}
have indicated that the proper-time region during which the
hydrodynamic flow is approximately (1+1) dimensional has a deep
impact on the whole process. Our aim is to take benefit of the
explicit form (\ref{Lansol3}) representative of  the entropy of
evolution-dominated flows, based on the Khalatnikov potential
(\ref{chilan}), to revisit the discussion in the light of recent
experimental results.

For the phenomenological application, we will concentrate on the entropy flow
corresponding to the Belenkij-Landau solution.
From \eqref{Lansol3} and \eqref{GentotSLan}, one obtains the
formula
\begin{equation}
\frac{dS}{dy}(y) = S_{tot}\ \frac{(g-1)}{4\sqrt g}\
e^{\frac{(g-1)}{2}\theta_F}
\left[I_0\left(\frac{g\!-\!1}{2}\sqrt{\theta_F^2-y^2/g}\right)-
I_1\left(\frac{g\!-\!1}{2}\sqrt{\theta_F^2-y^2/g}\right)
\frac{\theta_F}{\sqrt{\theta_F^2-y^2/g}}\right]\label{Lansol4}\ ,
\end{equation}
where we have used the normalization by the total entropy
$S_{tot}$. Formula (\ref{Lansol4}) still depends on  two
hydrodynamical parameters  $\theta_F,$ the logarithmic temperature
evolution, and the speed of sound $c_s=g^{-1/2}.$

\subsection{Multiplicity distribution at fixed energy}

In order to investigate the phenomenological validity of formula
(\ref{Lansol4}), let us consider the experimental BRAHMS data for
the charged multiplicity distribution in the most central
collisions as a function of the rapidity measured recently at RHIC
\cite{BRAHMS}. For sake of simplicity, in accordance with the
(1+1) dimensional approximation of the dynamics that we consider, we will make 
the following assumptions. We will identify the rapidity if the fluid elements 
$y_f\equiv 1/2 \log (u^+/u^-)$ with the rapidity of the particles $y_p\equiv 1/2 
\log(p^+/p^-)$. We thus keep the same notation $y$. In the same way, we assume 
that the
multiplicity distribution of produced particles ${dN}/{dy}$ in
rapidity can be considered  to be equal, up to a constant factor,
to the entropy distribution\footnote{We also assume
that the multiplicity distribution of charged particles is
proportional to the total one.} ${dS}/{dy}$.  One expects that the end of the
hydrodynamic behavior appears at a typical temperature $T_F$,
related to a hadronization or freeze-out temperature, and
independent of the total c.o.m. energy of the collision. On the
other hand, the initial temperature $T_0$ is expected to depend on
the total c.o.m. energy (or equivalently on the total rapidity
$Y$) and on the centrality of the  collision, through the energy
density $\epsilon(T_0)$ of the medium produced by the
pre-hydrodynamic stage of the collision. Thus, $\th_F=\log
T_F/T_0,$ should be a function of $Y$ and of the centrality. Our formalism, 
based on the $1+1$ dimensional approximation of the flow, is not appropriate to 
have a precise description of the freeze-out.
Note however that some improvement could be obtained
by using, $e.g.$, the Cooper-Frye formalism \cite{CF} in the
derivation of the entropy flow. We postpone this to further
studies.

\begin{figure}[ht]
\begin{center}
\mbox{\epsfig{figure=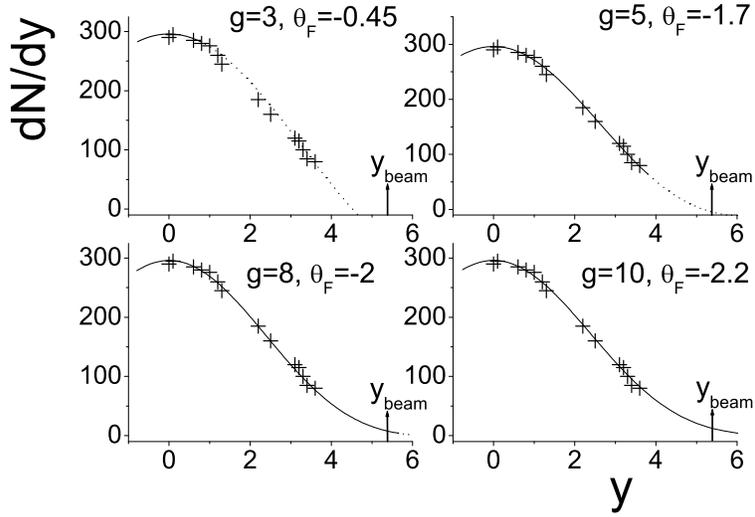,width=10cm,angle=0}}
\caption{{\it BRAHMS data fitted with the hydrodynamic formula.}
The data are taken from \cite{BRAHMS} and they correspond to
charged pions in cental Au+Au collisions at $\sqrt{s_{NN}}=200$
GeV. The solid line corresponds to the physical region
($\frac{y^2}{g}\le \theta_F^2$), while the dotted one corresponds
to its analytic continuation ($\frac{y^2}{g}>\theta_F^2$). The
small vertical line marks the experimental beam-rapidity
$Y_{beam}\sim Y/2$.} \label{datafit}
\end{center}
\end{figure}

Using then formula (\ref{Lansol4}) for ${dN}/{dy}$ and fitting
BRAHMS data by adjusting the parameter $\th_F$ we obtain a good
description for different values of $g$.  In Fig.~\ref{datafit},
as an example, we present the BRAHMS data  fitted with
(\ref{Lansol4}), for four pairs of $g$ and $\theta_F$ values,
reported on the figure.
In these plots the solid line corresponds to the physically
meaningful region ($\frac{y^2}{g}\le \theta_F^2$), while the
dotted line corresponds to the analytic continuation of formula
(\ref{Lansol4}) in the region $\frac{y^2}{g}>\theta_F^2,$  where
the applicability of the solution \eqref{Lansol4} is theoretically
questionable.

The phenomenological application appears to be correct for quite
different values of the speed of sound $c_s \equiv  g^{-1/2}.$ The
overall form of the curves is satisfactory. For the first curve at
$c_s = 1/\sqrt 3$ (\emph{i.e.} the conformal case), however,the  analytic
continuation beyond
$\frac{y^2}{g}\le \theta_F^2$ is soon reached\footnote{One may
also note that the curve indicates a violation of positivity
before the kinematical limit.}. We will comment on this remark
later on. Indeed,  when decreasing the speed of sound, $e.g.$ for
$g=5,$ the physical domain $\frac{y^2}{g}\le \theta_F^2$ extends
in rapidity.

Some comments on these results are in order.

a) It has been well-known since long \cite{Carruthers} and
confirmed more recently that a Gaussian fit to the data
\begin{equation}
\frac{dS}{dy}(y) \sim e^{-{y^2}/{Y}}\label{Lan1}
\end{equation}
with a variance $\sqrt Y$, as predicted by Landau \cite{lan}, was
reasonably verified. We noticed, that, indeed, expression
(\ref{Lansol4}) has an approximate Gaussian form, but it does not
correspond, except for very large $\th_F,$ to the expansion  of
the exact entropy distribution near $y=0$, as in the original
argument \cite{lan} which was  based on an asymptotic
approximation\footnote{It is indeed easy to verify that for
phenomenological values of  $\theta_F$, this approximation does
not work in the data range.}. Hence, the sub-asymptotic features
of the full solution plays an important phenomenological role.

b) There is  apparently no track  of the transition between
the physical regime $\frac{y^2}{g}\le \theta_F^2$ and its analytic
continuation, described  by the dotted lines in Fig.\ref{datafit}.
This is related to the mathematical property of the general
solution (\ref{genentr}) expressed using  a Laplace transform. In
short, the $I_{0,1}$ Bessel functions are transformed into
$J_{0,1}$ with the same argument up to a factor $i$, without
discontinuity.

c) This transition is however meaningful. In fact one knows that
the lines $\frac{y}{\sqrt g}= \mp \theta_F$ delineate different
regions of the hydrodynamical regime. Discontinuities, and thus
shock or Riemann waves may occur at these boundaries, called
characteristics of the equation \cite{Hilbert}. Hence some other
solutions may branch at this point (see $e.g.$ \cite{bi,old}).
However, our results do not depend on the specific form of these
other solutions such as the shock waves considered in
\cite{bi,old}.

\subsection{Energy dependence of the multiplicity distributions}
Going a step further, we would like to interpret the energy
dependence ($i.e.$ the $Y$ dependence)  of the (1+1) solution for
the entropy flow compared with multiplicity data. For this sake,
we make use of the Gaussian fits reported\footnote{In fact, the
prediction  (\ref{Lan1}) fits reasonably well, but we used instead
the actual best-fit determination of the variances  provided in
\cite{BRAHMS}.} in  \cite{BRAHMS} for different sets of data
ranging from the AGS to RHIC.

\begin{figure}[ht]
\begin{center}
\mbox{\epsfig{figure=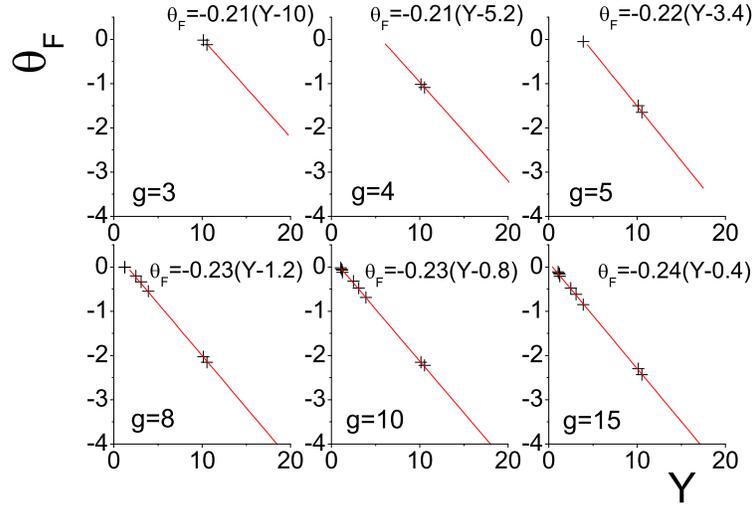,width=10cm,angle=0}} \caption{{\it
The hydrodynamical parameter $\theta_F$ as a function of $Y.$} We
describe the dependence for for six values of $g,$ in a large
range starting from the canonical value $3$ which would correspond
to a conformal fluid in $(3+1)$ dimensions.} \label{thetaL}
\end{center}
\end{figure}

In Fig.~\ref{thetaL} we give the determination of $\theta_F$ as a
function of   $Y$ which gives a good description of the  Gaussian
fits with the variance taken from \cite{BRAHMS}. As in the
previous study of BRAHMS data, we performed this fit for six
different values of $g$. As shown in Fig.~\ref{thetaL} the
corresponding relation is clearly linear.  We can write:
\begin{equation}
\theta_F=-\kappa\left(Y-Y_0\right)\label{thetaLL}\ .
\end{equation} As shown on the figure, the constant term $Y_0$
depends appreciably on $g$ while  the slope $\kappa\approx 0.2$
remains only slightly dependent on it.
On a physical ground, the linear relation (\ref{thetaLL}) has a
reasonable interpretation.  The initial temperature of the medium
is expected to grow as a power $\kappa\le 1$ of the incident
energy.
 One finds approximately
$T_0/T_F = e^{-\th_F} \sim e^{.22 (Y-Y_0)}.$ Hence the more energy
is available, the longer the hydrodynamical evolution lasts.  At
smaller speed of sound the hydrodynamical evolution has to occur
on a larger temperature interval in order to describe the same
entropy distribution, as could be expected.

Moreover, there is a physical argument, analogous to the one
proposed by Landau \cite{lan}, for the existence of a linear
relation (\ref{thetaLL}) between the temperature ratio and the
total c.o.m. energy. Assuming the approximate validity of the
Bjorken relation\footnote{This relation, properly stating, is
exact only for the Bjorken boost-invariant flow. However, one
expects that it remains approximately valid  in the central region
of more general flows (see, $e.g.$ \cite{Bipe}.)} $\tau_0/\tau_F
\approx \left(T_F/T_0\right)^g,$ where $\tau_0$ (resp. $\tau_F$)
are the initial (resp. final) proper-times of the (1+1)
hydrodynamical evolution and reporting in (\ref{thetaLL}) one
finds
\begin{equation}
\log \tau_0/\tau_F \approx \kappa g
\left(Y-Y_0\right)\label{tauLL}\ .
\end{equation}
Indeed, following \cite{lan}, the separation proper-time from
(1+1) hydrodynamics  to the (1+3) regime is of order $\log
\Delta\tau_s \sim \f12 Y$ where $\Delta$ is the typical transverse
size of the initial particles. Assuming that we can approximate
$\Delta\tau_s$ by $\tau_0/\tau_F,$ and taking into account the
Bjorken flow approximation,  formula (\ref{tauLL}) is suggestive
of the proportionality property. We leave the precise values for
$\kappa$, $\tau_0/\tau_F$ and $Y_0$ to a further determination of
$g$ since the data we discussed do not prefer a precise value of
$\kappa g$ (to be compared with $1/2$ obtained for $\log
\Delta\tau_s$).

An interesting consequence of the linear relation (\ref{thetaLL})
between  $\th_F$ and $Y$ at fixed $g$ is the possibility of
relating the general hydrodynamic entropy distribution
(\ref{genentr}) to the one-particle inclusive cross-section and
thus to the appropriate scattering amplitudes. These are not easy
to formulate  in the hydrodynamical formalism. Being more
specific, let us transform the formula (\ref{genentr}) in terms of
the energy  dependence using  $\kappa$ as the coefficient of
proportionality in (\ref{thetaLL}). We get

\begin{eqnarray}
\frac{dN}{dy}(y,Y) \propto \
\int^{\gamma_0+i\infty}_{\gamma_0-i\infty}\frac{d\gamma} {2\pi i}\
e^{-\kappa (Y-Y_0)(\gamma+\frac{g-1}{2})}\left[\left(\gamma\!+\!
g/2\right)^2\!-\!1/4\right]\  {\tilde{K}(\gamma)}\ \frac
{e^{{-\frac{|y|}{\sqrt{g}}\sqrt{\gamma^2-
\frac{(g-1)^2}{4}}}}}{\sqrt{\gamma^2-\frac{(g-1)^2}{4}}}\ \,\
.\label{dsdy2}
\end{eqnarray}
Formula \eqref{dsdy2} shows that the characteristic
hydrodynamic  kernel $\cal Q$ (see \eqref{kernel}) appears also as
the kernel of the Laplace transform in $Y$ of the one-particle
inclusive cross-section, up to a redefinition of the conjugate
moment $\om=\kappa\g$ of the total rapidity $Y.$ This relation
may be useful to compare various theories and models for
scattering amplitudes of high-energy collisions with the
predictions of hydrodynamic evolution.

\section{Conclusions and outlook}
Let us summarize the results of our study:

On the  theoretical ground, we have the following results:

i) We have recalled and reformulated the derivation of the Khalatnikov potential
and equation in terms of light-cone variables.  This allows to formulate the
initial nonlinear problem of (1+1) hydrodynamics of a perfect fluid in terms of
solutions of a linear equation. As an application, using the Green function
formalism we derive the general form of the solution for constant speed of
sound.

ii) Expressing the flow of entropy through fixed-temperature hypersurfaces, we
provide
general and simple expressions  of the entropy flow
${dS}/{dy}$ in terms of the Khalatnikov potential\footnote{After the completion 
of this paper, we noticed a related study in Ref.\cite{milekhin}}.

iii) We check and illustrate the simplicity of the obtained formulae for $dS/dy$
by applying the formalism to some exact hydrodynamic solutions which were not
using the Khalatnikov formulation, such as the Bjorken flow and the less
straightforward example of the harmonic flows of Ref.\cite{Bipe}.

iv) We use our formalism to find the entropy flow for the subclass of solutions
for which the hydrodynamic evolution dominates over the infliuence of initial
conditions. A characteristic example  of such flows is the one studied long ago
by Landau and Belenkij \cite{bi}, corresponding to full stopping initial
conditions. We provide an exact expression for the related entropy flow.

As a phenomenological application, we discuss  the relevance of the full
stopping entropy flow for modern heavy-ion experiments which was advocated,
\emph{e.g.} in Refs.\cite{Carruthers,stein}.

i) The exact expression of ${dS}/{dy}$ for the Belenkij-Landau
solution, depending only on the ratio $T_F/T_0$ and on the speed
of sound $c_s$, is in agreement with the shape of the multiplicity
distribution of particles $ {dN}/{dy}(y,Y)$ observed in heavy-ion
reactions, with a linear relation between the temperature ratio
and the total rapidity $\log T_0/T_F = \kappa \left[Y -Y_0(c_s)\right].$

ii) However, comparing our exact results with the asymptotic Gaussian
predictions \cite{lan,Carruthers} for the multiplicity distributions, we find
that  nonasymptotic contributions play an important role in the phenomenological
description.

iii) The speed of sound, which is the remaining parameter in our
study, is not determined by the multiplicity distribution, since
the phenomenological description seems satisfactory for a rather
large range of the parameter $g\equiv 1/c_s^2.$ However, even if
one does not see any sizable effect on the curve for
${dS}/{dy}(y,T)$ one notices that the physical domain of the
hydrodynamic expansion is restricted by the condition ${y^2}\le g\
\theta_F^2$, especially for a speed of sound as large as the
 conformal one $c_s=1/\sqrt{3}.$

This summary of conclusions leads to a few comments on possible
further  developments of our approach.  Some of them are technical
but could provide a further insight on the features of (1+1)
hydrodynamics. First, it should be useful to study in detail a
larger set of solutions.
  Second, implementing the Cooper-Frye
formalism \cite{CF} directly in terms of the Khalatnikov potential
could refine the hypothesis of a fixed final temperature $T_F.$
Also, the investigation of the entropy flow through other
hypersurfaces, in particular the proper-time ones ($cf.$
\cite{Bipe}) would be welcome, in particular to allow for a
straightforward implementation of fixed proper-time initial
conditions.

On the phenomenological point-of-view, it is important to develop
the  comparison of the (1+1) approach with the data and including
more corrections to the idealized dominance of the longitudinal
motion. One question could be settled at least phenomenologically,
which is the determination of the best fit for the speed of sound
parameter, which is presently rather free. Also, including a viscosity
contribution is another important issue, together with the
investigation of the entropy flows with varying speed-of-sound.

One may ask what is the
meaning of the transition on the lines ${y^2}={g}\theta_F^2,$
which appear even in the physical rapidity region. Mathematically,
they are the Riemann characteristics of the Khalatnikov equations
and as such, they are regions where discontinuities may appear
\cite{Hilbert}. Indeed, these characteristics were used in the old
studies \cite{bi,old} to connect boundary Riemann waves to the
domain of dynamical hydrodynamic evolution. What is their meaning,
if any,  in the modern view we have now of high-energy collisions,
is an interesting open question.

On a more conceptual point-of-view, our study of the entropy
flow and its dependence on rapidity may have some impact on recent
studies \cite{janik,autres} of the AdS/CFT correspondence. It
relates  the hydrodynamics of a fluid, whose microscopic
description is the one of a gauge field theory, with the string
theory in a higher-dimensional space where the Einstein equations
govern the gravitational properties of its low-energy regime. The
actual realizations of the duality correspondence for a collective
flow require boost-invariance and thus are limited to the Bjorken
flow. This flow contains an infinite energy, and is thus of
limited relevance. Knowing the analytic form of more physical
solutions should be helpful to derive their dual gravitational
backgrounds, which is \emph{a priori} a formidable task.
\\

\paragraph*{{\bf{Acknowledgements:}}}
We thank Andrzej Bialas and Jean-Yves Ollitrault for useful suggestions.
One of us (E.N.S) wishes to thanks Institut de Physique
Th\'eorique, CEA,  for the hospitality during the preparation of
the present work.

\end{document}